
\let\dvips=\relax       


\ifx\files\undefined \input indexes \fi

\files


\ifx\loadfont\undefined \input fonts \fi
\xiifonts \xiititles \rm
 
\out{tex}\utfon

\font\sfit=cmssi12

\catcode`\@=11

\font\xiiex=cmex10 at12pt
\textfont3=\xiiex \scriptfont3=\xiiex \scriptscriptfont3=\xiiex

\font\xiibb=msbm10 at12pt
\font\ixbb=msbm9
\font\vibb=msbm6
\newfam\bbfam 
\textfont\bbfam=\xiibb \scriptfont\bbfam=\ixbb
\scriptscriptfont\bbfam=\vibb

\mathchardef\subsetneq="3828

\font\xiifrak=eufm10 at12pt
\font\ixfrak=eufm9
\newfam\frakfam \textfont\frakfam=\xiifrak \scriptfont\frakfam=\ixfrak

\mathcode`?="603F 
\def\ifmath$#1${\relax\ifmmode #1\else$#1$\fi}


\ifcase\pdfoutput
 
\else
 \pdfcompresslevel=9
 \pdfdecimaldigits=4
 \pdfhorigin=1truein
 \pdfvorigin=1truein
 \pdfpageheight=297truemm
 \pdfpagewidth=210truemm
\fi
\hsize=15.92truecm \vsize=24.62truecm 
\advance\vsize -30pt

\def\twodigits#1{\ifnum #1<10 0\fi \number#1}

\let\version=\todayiso
\def\Folio{\ifnum\pageno<0
 \uppercase\expandafter{\romannumeral-\pageno}\else\number\pageno\fi}

\headline={\vrule height10.2pt depth4.2pt width0pt
 {\xiitt www.ramoncasares.com}\quad{\xiirm\version}\hfil
 \quad{\xiibf\jobname\quad\Folio}}%
\def\makeheadline{\vbox to 30pt{\colorblack\line{\the\headline}%
  \kern 1pt \hrule height 1pt\vfil\endcolor}\nointerlineskip}
\nopagenumbers


\newcount\secno
\newcount\ssecno
\newcount\thno
\newcount\parno
\let\presec=\empty
\def\presec{^^c2^^a7}

\parskip=0pt plus 1pt
\newdimen\oldparindent \oldparindent=20pt \parindent=0pt
\def\hang{\hangindent\oldparindent}

\def\numberedpars{\global\advance\parno1 
 \noindent\hbox to\oldparindent{{\xiiscriptsy\char123
  \xiiscriptrm\number\parno}\hfil$\cdot$\hfil}\ignorespaces}
\def\continuepar{{\everypar{}\par\noindent}\hangindent0pt\ignorespaces}

\def\begincenter{\par\begingroup \parindent=0pt
 \advance\leftskip 0pt plus 1fill \advance\rightskip 0pt plus 1fill
 \def\\{\unskip\break\ignorespaces}}
\def\endcenter{\par\endgroup}

\let\n@xtlbl=\relax
\def\labeled#1{\def\n@xtlbl{#1}}

\outer\def\section#1{\vskip0pt plus 90pt\penalty-500\vskip0pt plus-90pt
 \goodbreak\vskip 2pc plus 1pc minus 3pt
 \ifx\n@xtlbl\relax \def\1{#1}\else\let\1=\n@xtlbl \let\n@xtlbl=\relax \fi
 \everypar{}\advance\secno1
 \ssecno=0 \thno=0 \parno=0
 \def\secid{\presec\the\secno}\noindent\dest
 {\setbox2=\hbox{\fonttwo\secid\enspace}\hangindent\wd2
  \rightskip=0pt plus 6em
  \fonttwo\secid\enspace#1\toc1{#1}\lbl{\1}{#1}\par}
 \everypar{\numberedpars}}

\outer\def\xsection#1{\vskip0pt plus 90pt\penalty-500\vskip0pt plus-90pt
 \goodbreak\vskip 2pc plus 1pc
 \ifx\n@xtlbl\relax \def\1{#1}\else\let\1=\n@xtlbl \let\n@xtlbl=\relax \fi
 \everypar{}\parno=0
 \def\secid{}\noindent\dest
 {\rightskip=0pt plus 6em
  \fonttwo#1\toc1{#1}\lbl{\1}{#1}\par}}

\outer\def\subsection#1{\vskip0pt plus 60pt\penalty-250\vskip0pt plus-60pt
 \ifnum\parno=0 \vskip 0.5pc plus 3pt minus 3pt\else\vskip 1pc plus 6pt minus 6pt\fi
 \ifx\n@xtlbl\relax \def\1{#1}\else\let\1=\n@xtlbl \let\n@xtlbl=\relax \fi
 \everypar{}\advance\ssecno1 \thno=0 \parno=0
 \def\secid{\presec\the\secno.\the\ssecno}\noindent\dest
 {\setbox2=\hbox{\fontthree\secid\enspace}\hangindent\wd2
  \rightskip=0pt plus 6em
 \fontthree\secid\enspace#1\toc2{#1}\lbl{\1}{#1}\par}
 \everypar{\numberedpars}}

\outer\def\xsubsection#1{\vskip0pt plus 60pt\penalty-250\vskip0pt plus-60pt
 \ifnum\parno=0 \vskip 0.5pc plus 3pt minus 3pt\else\vskip 1pc plus 6pt minus 6pt\fi
 \ifx\n@xtlbl\relax \def\1{#1}\else\let\1=\n@xtlbl \let\n@xtlbl=\relax \fi
 \toks0=\everypar \everypar{}\def\secid{}\noindent\dest
 {\rightskip=0pt plus 6em
  \fontthree#1\toc2{#1}\lbl{\1}{#1}\par}
 \everypar=\toks0}

\outer\def\clause#1{\vskip0pt plus 30pt\penalty-150\vskip0pt plus-30pt
 \everypar{}\medskip\hang\noindent
 \advance\thno1 \def\secid{\presec\the\secno.\the\ssecno.\the\thno}%
 {\sc\secid\ #1}\quad\ignorespaces}


\def\title#1{\def\titledoc{#1}}
\def\author#1{\def\authordoc{#1}}
\def\keywords#1{\def\keywordsdoc{#1}}
\def\subject#1{\def\subjectdoc{#1}}
\def\contact#1{\def\contactdoc{#1}}

\def\beginnote{\insert\footins\bgroup\continuepar\strut\colorblack}
\def\endnote{\endcolor\egroup}

\def\infodoc{\begingroup
  \def~{ }\def\\{ }\def\S{^^a7}\let\-\relax \stringaccents
  \deactivatehigh \utfon\out{pdf}\let^^e2=\@eii \doextrapdf
  \ifcase\pdfoutput
   \ifx\dvips\undefined
   \else
    \special{ps: [
     /Title (\titledoc) /Author (\authordoc)
     /Keywords (\keywordsdoc) /Subject (\subjectdoc)
     /DOCINFO pdfmark }%
   \fi
  \else
   \pdfinfo{/Title (\titledoc) /Author (\authordoc)
     /Keywords (\keywordsdoc) /Subject (\subjectdoc)}\fi
 \endgroup}

\def\maketitle{\par \infodoc
 \null\vskip 3pc plus 4pc minus 1pc
 \def\secid{}\dest\toc0{\titledoc}\lbl{\titledoc}{}
 \begincenter \fontzero\titledoc \endcenter
 \vskip 1pc plus 1pc
 \centerline{\fontone\authordoc}
 \ifx\contactdoc\undefined\else
  \vskip 6pt minus 3pt
  \centerline{\rm\contactdoc}\fi
 \vskip 2pc plus 1pc minus 1pc\relax}

\def\beginabstract{\begingroup \parindent=55pt\narrower \sl\parindent=20pt\noindent}
\def\endabstract{\par \ifx\keywordsdoc\undefined\else \sfit\rightskip=55pt plus3em
 \smallskip\setbox0=\hbox{Keywords:\quad}\hangindent\wd0
 \noindent\box0\keywordsdoc\par\fi \smallskip\endgroup}



\def\citenone #1 (#2){\ref{#1#2}}
\def\citetext [#1] #2 (#3){\refx{{\def\1{#2}\def\2{#3}#1}}{#2#3}}
\def\citestar * #1 (#2){\refx{#1\ #2}{#1#2}}

\def\cite{\futurelet\nexttoken\citexx}
\def\citexx{\if *\nexttoken \let\next=\citestar \else
 \if [\nexttoken \let\next=\citetext \else
 \let\next=\citenone \fi\fi\next}

\def\chreference #1 (#2){\everypar{}\par
 \vskip0pt plus 2\baselineskip\penalty-43
 \vskip0pt plus-2\baselineskip
 \noindent\hangindent20pt\relax
 #1\thinspace(#2)\dest\lbl{#1#2}{#1\ (#2)}}

\def\brreference #1 (#2){\everypar{}\par
 \vskip0pt plus 2\baselineskip\penalty-43
 \vskip0pt plus-2\baselineskip
 \noindent\hangindent20pt\relax
 [#1 #2]\dest\lbl{#1#2}{[#1 #2]}}

\let\reference=\chreference
\let\biblabel=\chreference

\def\references{\vskip0pt plus 90pt\penalty-500\vskip0pt plus-90pt
 \everypar{}\parno=0
 \goodbreak\vskip 2pc plus 1pc
 \def\secid{}
 \noindent{\fonttwo References\dest\toc1{References}\lbl{References}{}}\par
 \begingroup
  \catcode`\_=13
  \def\biblabel ##1 ({{\stringaccents\def\&{\string\&}\xdef\1{##1}}\next}
  \def\next ##1) ##2\par{
    \expandafter\gdef\csname \1##1\endcsname{##2}}

\hyphenation{pro-p-o-si-tions Ent-schei-dungs-prob-lem}

\biblabel Abelson \& Sussman (1985)
Harold Abelson, and Gerald Sussman, with Julie Sussman,
\book{Structure and In\-terpretation of Computer Programs};
The MIT Press, Cambridge MA, 1985,
\ISBN 978-0-262-01077-1.

\biblabel ASCII (1963)
\article{American Standard Code for Information Interchange};
ASA X3.4-1963,
American Standards Association, New York, NY, 1963.

\biblabel Barendregt (1985)
Henk P.\ Barendregt,
\book{The Lambda Calculus, its Syntax and Semantics};
Revised Edition, Studies in Logic and the Foundations of Mathematics,
Vol.\ 103, North-Holland Publishing Co, Amsterdam, 1985,
\ISBN 0-444-87508-5.

\biblabel Bell (1964)
John Bell,
\article{On the Einstein Podolsky Rosen Paradox};
in \periodical{Physics},
vol.\ 1, pp.\ 195--200, 1964,
\DOI{10.1142/9789812386540_0002}.

\biblabel Berwick \& Chomsky (2016)
Robert C.\ Berwick, and Noam Chomsky,
\book{Why Only Us: Language and Evolution};
The MIT Press, Cambridge MA, 2016,
\ISBN 978-0-262-03424-1.

\biblabel Bickerton (1990)
Derek Bickerton,
\book{Language \& Species};
The University of Chicago Press, Chicago, 1990,
\ISBN 978-0-226-04611-2.

\biblabel Bickerton (2009)
Derek Bickerton,
\book{Adam's Tongue: How Humans Made Language, How Language Made Humans};
Hill and Wang, New York, 2009,
\ISBN 978-0-8090-1647-1.

\biblabel Bickerton (2014)
Derek Bickerton,
\book{More than Nature Needs: Language, Mind, and Evolution};
Harvard University Press, Cambridge MA, 2014,
\ISBN 978-0-674-72490-7.

\biblabel Bohm (1952)
David Bohm,
\article{A Suggested Interpretation of the Quantum Theory
in Terms of ‘Hidden’ Variables}; two parts
in \periodical{Physical Review},
vol.\ 85, pp.\ 166--179 and 180--193, January 1952,
\DOI{10.1103/PhysRev.85.166} and\\
\DOI{10.1103/PhysRev.85.180}.

\biblabel Bruiger (2017)
Dan Bruiger,
\article{Physics and Fundamentalism:
Science as the Continuation of Religion by Other Means};
{\tt \URL <https://www.academia.edu/33484236>}.

\biblabel Button (2009)
Tim Button,
\article{SAD Computers and Two Versions
of the Church-Turing Thesis};
in \periodical{The British Journal for the Philosophy of Science},
vol.\ 60, no.\ 4, pp.\ 765--792, December 2009,
\DOI{10.1093/bjps/axp038}.

\biblabel Casares (B)
Ramón Casares,
\article{Biolinguistics XXI: Semantics and Pragmatics};\\
\DOI{10.6084/m9.figshare.11300558}.

\biblabel Casares (C)
Ramón Casares,
\article{Proof of Church's Thesis};\\
\DOI{10.6084/m9.figshare.4955501},
{\tt\URL arXiv:1209.5036<http://arxiv.org/abs/1209.5036>}.

\biblabel Casares (E)
Ramón Casares,
\article{Errors in Infinite Computations};
\DOI{10.6084/m9.figshare.13686439}.

\biblabel Casares (H)
Ramón Casares,
\article{A Complete Hierarchy of Languages};\\
\DOI{10.6084/m9.figshare.6126917}.

\biblabel Casares (I)
Ramón Casares, \article{The Intention of Intention};\\
\DOI{10.6084/m9.figshare.7928240}.

\biblabel Casares (J)
Ramón Casares,
\article{Putnam's Rocks Are Clocks};\\
\DOI{10.6084/m9.figshare.5450278}.

\biblabel Casares (K)
Ramón Casares,
\article{Subjectivist Propaganda};\\
\DOI{10.6084/m9.figshare.13076906}.

\biblabel Casares (M)
Ramón Casares,
\article{Merge Is Not Recursion};\\
\DOI{10.6084/m9.figshare.4958822}.

\biblabel Casares (P)
Ramón Casares,
\article{Problem Theory};\\
\DOI{10.6084/m9.figshare.4956353},
{\tt\URL arXiv:1412.1044<http://arxiv.org/abs/1412.1044>}.

\biblabel Casares (R)
Ramón Casares,
\article{On ‘On Recursion’\thinspace};\\
\DOI{10.6084/m9.figshare.5097691}.

\biblabel Casares (S)
Ramón Casares,
\article{Syntax Evolution: Problems and Recursion};\\
\DOI{10.6084/m9.figshare.4956359},
{\tt\URL arXiv:1508.03040<http://arxiv.org/abs/1508.03040>}.

\biblabel Casares (T)
Ramón Casares,
\article{On Turing Completeness, or Why We Are So Many};\\
\DOI{10.6084/m9.figshare.5631922}.

\biblabel Casares (U)
Ramón Casares,
\article{Universal Grammar is a universal grammar};\\
\DOI{10.6084/m9.figshare.4956764}.

\biblabel Chomsky (1957)
Noam Chomsky,
\book{Syntactic Structures};
The Hague, Mouton \& Co., 1957, 2002.
\ISBN 3-11-017279-8.

\biblabel Chomsky (1959)
Noam Chomsky,
\article{On Certain Formal Properties of Grammars};\\
in \periodical{Information and Control},
vol.\ 2, no.\ 2, pp.\ 137--167, June 1959,\\
\DOI{10.1016/S0019-9958(59)90362-6}.

\biblabel Chomsky (1965)
Noam Chomsky,
\book{Aspects of the Theory of Syntax};
The MIT Press, Cambridge MA, 1965,
\ISBN 978-0-262-53007-1.

\biblabel Chomsky (1986)
Noam Chomsky,
\book{Knowledge of Language: Its Nature, Origin, and Use},
Convergence series;
Praeger, New York, 1986,
\ISBN 978-0-275-91761-6.

\biblabel Chomsky (1988)
Noam Chomsky,
\book{Language and Problems of Knowledge: The Managua Lectures},
Current Studies in Linguistics, 16;
The MIT Press, Cambridge MA, 1988,
\ISBN 978-0-262-53070-5.

\biblabel Chomsky (2000)
Noam Chomsky,
\book{New Horizons in the Study of Language and Mind};
Cambridge University Press, Cambridge, 2000,
\ISBN 978-0-521-65822-5.

\biblabel Chomsky (2005a)
Noam Chomsky,
\article{Some Simple Evo-Devo Theses:
 How True Might They Be for Language?};
Paper of the talk given to the
Morris Symposium on the Evolution of Language,
held at Stony Brook University, New York,
in October 15, 2005;
{\sc url:}
\URL<https://linguistics.stonybrook.edu/events/morris/05/program>.

\biblabel Chomsky (2005)
Noam Chomsky,
\article{Three Factors in Language Design};
in \periodical{Linguistic Inquiry},
vol.\ 36, no.\ 1, pp.\ 1--22, Winter 2005,
\DOI{10.1162/0024389052993655}.

\biblabel Chomsky (2006)
Noam Chomsky,
\book{Language and Mind}, Third Edition;
Cambridge University Press, Cambridge, 2006,
\ISBN 978-0-521-67493-5.

\biblabel Chomsky (2007)
Noam Chomsky,
\article{Of Minds and Language};
in \periodical{Biolinguistics},
vol.\ 1, pp.\ 9--27, 2007,\\
{\sc url:}
\URL<http://www.biolinguistics.eu/index.php/biolinguistics/article/view/19>.


\biblabel Church (1935)
Alonzo Church,
\article{An Unsolvable Problem of Elementary Number Theory};
in \periodical{American Journal of Mathematics},
vol.\ 58, no.\ 2, pp.\ 345--363, April 1936,
\DOI{10.2307/2371045}.
Presented to the American Mathematical Society,
April 19, 1935.

\biblabel Church (1937)
Alonzo Church,
\article{Review of \cite Post (1936)};
in \periodical{The Journal of Symbolic Logic},
Volume 2, Issue 1, p.~43, March 1937.
\DOI{10.1017/S0022481200039591}.

\biblabel Curry \& Feys (1958)
Haskell B.\ Curry, and
Robert Feys, with
William  Craig,
\book{Combinatory Logic}, Vol.\ I;
North-Holland, Amsterdam, 1958,
\ISBN 978-0-7204-2207-8.

\biblabel Darwin (1859)
Charles Darwin,
\book{On the Origin of Species by Means of Natural Selection,
  or the Preservation of Favoured Races in the Struggle for Life};
John Murray, London, 1859,
\URL Project Gutenberg<https://archive.org/download/rmcg0005/Darwin-OriginOfSpecies-a1.pdf>.

\biblabel Davies \& Brown (1986)
Paul Davies and Julian R. Brown (editors),
\book{The Ghost in the Atom:
 A Discussion of the Mysteries of Quantum Physics};
Cambridge University Press, Cambridge, 1986,
\ISBN 0-521-45728-9.

\biblabel Davis (1965)
Martin Davis (editor),
\book{The Undecidable: Basic Papers on Undecidable Propositions,
Unsolvable Problems and Computable Functions};
Dover, Mineola, New York, 2004,
\ISBN 978-0-486-43228-1.
Corrected republication of the same title
by Raven, Hewlett, New York, 1965.

\biblabel Davis (1982)
Martin Davis,
\article{Why Gödel Didn't Have Church's Thesis};
in \periodical{Information and Control},
vol.\ 54, pp.\ 3--24, 1982,
\DOI{10.1016/s0019-9958(82)91226-8}.

\biblabel Descartes (1637)
René Descartes,
\book{Discourse on the Method
 of Rightly Conducting One's Reason
 and of Seeking Truth in the Sciences};
translated by John Veitch (1850),
\URL Project Gutenberg<https://archive.org/download/rmcg0001/Descartes-Discourse-a1.pdf>.

\biblabel Descartes (1641)
René Descartes,
\book{Meditations on First Philosophy,
 in which the existence of God
 and the immortality of the soul are demonstrated};
translated by John Veitch (1852),
\URL Lancaster University<https://archive.org/download/RMCG0002/Descartes-Meditations-a1.pdf>.

\biblabel DeLong (1970)
Howard DeLong,
\book{A Profile of Mathematical Logic};
Dover, Mineola, New York, 1998,
\ISBN 0-486-43475-3.

\biblabel Deutsch (1985)
David Deutsch,
\article{Quantum theory, the Church-Turing principle and
  the universal quantum computer};
in \periodical{Proceedings of the Royal Society A},
 vol.\ 400, pp.\ 97--117, 1985,
\DOI{10.1098/rspa.1985.0070}.

\biblabel Dobzhansky (1973)
Theodosius Dobzhansky,
\article{Nothing in Biology Makes Sense
  Except in the Light of Evolution};
in \periodical{The American Biology Teacher},
 vol.\ 35, no.\ 3 (March, 1973), pp.\ 125--129,
\DOI{10.2307/4444260}.

\biblabel Everett (2008)
Daniel L.\ Everett,
\book{Don't Sleep, There Are Snakes:
Life and Language in the Amazonian Jungle};
Vintage, New York, 2008,
\ISBN 978-0-307-38612-0.

\biblabel Feyerabend (1988)
Paul Feyerabend,
\book{Against Method};
Revised Edition, Verso, London, 1988,
\ISBN 0-86091-934-X.

\biblabel  Fitch, Hauser, and Chomsky (2005)
Tecumseh Fitch, Marc Hauser, and Noam Chomsky,
\article{The Evolution of the Language Faculty:
Clarifications and Implications};
in \periodical{Cognition},
vol.\ 97, pp.\ 179--210, 2005,
\DOI{10.1016/j.cognition.2005.02.005}.

\biblabel Fortran (1956)
International Business Machines Corporation,
\book{The Fortran Automatic Coding System for the IBM 704 EDPM};
International Business Machines Corporation, New York,
Programmer's Reference Manual, October 15, 1956.

\biblabel Fortran (1966)
American National Standards Institute,
\book{FORTRAN: USAS X3.9-1966};
United States of America Standards Institute, New York, NY, 1966.

\biblabel Fortran (1977)
International Organization for Standardization,\\
\book{FORTRAN: ISO 1539:1980};
\URL<https://www.iso.org/standard/6127.html>.

\biblabel Fortran (1990)
International Organization for Standardization,\\
\book{Fortran: ISO/IEC 1539:1991};
\URL<https://www.iso.org/standard/17366.html>.

\biblabel Friedman \& Felleisen (1987)
Daniel Friedman, and Matthias Felleisen,
\book{The Little LISPer}, Trade Edition;
The MIT Press, Cambridge MA, 1987,
\ISBN 978-0-262-56038-2.

\biblabel von Frisch (1973)
Karl von Frisch,
\article{Decoding the Language of the Bee};
in \periodical{Science},
vol.\ 185, no.\ 4152, pp.\ 663--668, 23 Aug 1974,
\DOI{10.1126/science.185.4152.663}.
Nobel Lecture, December 12, 1973.

\biblabel Gandy (1980)
Robin Gandy,
\article{Church's Thesis and Principles for Mechanisms};
\DOI{10.1016/s0049-237x(08)71257-6}.
In \book{The Kleene Symposium}
(editors: J.\ Barwise, H.J.\ Keisler \& K.\ Kunen),
Volume 101 of
Studies in Logic and the Foundations of Mathematics;
North-Holland, Amsterdam, 1980, pp.~123--148;
\ISBN 0-444-85345-6.

\biblabel Ginsburg (2016)
Jason Ginsburg,
\article{Modeling of problems of projection:
 A non-counter\-cyclic approach};
in \periodical{Glossa},
vol.\ 1, no.\ 1, art.\ 7, pp.\ 1--46, 2016,
\DOI{10.5334/gjgl.22}.

\biblabel Gödel (1930)
Kurt Gödel,
\articlede{Über formal unentscheidbare Sätze der Principia Mathematica
 und verwandter Systeme I};
in \periodical{Monatshefte für Mathematik und Physik},
vol.\ 38, pp.\ 173--198, 1931,
\DOI{10.1007/BF01700692}.
Received November 17, 1930.
English translation in \cite Davis (1965).

\biblabel Gödel (1946)
Kurt Gödel,
\article{Remarks before the Priceton Bicentennial Conference on
  Problems in Mathematics};
in \cite Davis (1965), pp.\ 84--88.

\biblabel Gold (1967)
Mark Gold,
\article{Language Identification in the Limit};\\
in \periodical{Information and Control},
Volume 10, Issue 5, pp.~447--474, May 1967,\\
\DOI{10.1016/S0019-9958(67)91165-5}.

\biblabel Gregory (1988)
Bruce Gregory,
\book{Inventing Reality: Physics as Language};
John Wiley \& Sons, New York, 1988,
\ISBN 0-471-61388-6.

\biblabel Hamblin (1973)
Charles Hamblin,
\article{Questions in Montague English};
in \periodical{Foundations of Language},
vol.\ 10, pp.\ 41--53, 1973.

\biblabel Hauser, Chomsky, and Fitch (2002)
Marc Hauser, Noam Chomsky, and Tecumseh Fitch,
\article{The Language Faculty:
 Who Has It, What Is It, and How Did It Evolved?};
in \periodical{Science} 298, pp.\ 1569--1579, 2002,
\DOI{10.1126/science.298.5598.1569}.

\biblabel Hilbert (1900)
David Hilbert.
\article{Mathematical Problems};
Lecture delivered before the international
congress of mathematicians at Paris in 1900.
In \periodical{Bulletin of the American Mathematical Society},
vol.\ 8, no.\ 10, pp.\ 437--479, July 1902.
Translated by Dr.\ Mary Winston Newson.
\DOI{10.1090/S0002-9904-1902-00923-3}.

\biblabel Hilbert (1922)
David Hilbert,
\articlede{Neubegründung der Mathematik: Erste Mitteilung};
in \periodical{Abhandlungen aus dem Seminar
 der Hamburgischen Universität},
Volume 1, Issue 1, pp.~157--177, December 1922;
\DOI{10.1007/bf02940589}.
Spanish translation by L.F.\ Segura as
\article{La Nueva Fundamentación de las Matemáticas},
in \book{Fundamentos de las Matemáticas},
 (editors: C.\ Álvarez \& L.F.\ Segura),
Servicios Editoriales de la Facultad de Ciencias, UNAM,
México, 1993, pp.~37--62;
\ISBN 968-36-3275-0.
English translation by W.B.\ Ewald as
\article{The New Grounding of Mathematics: First Report},
in \book{From Kant to Hilbert:
 A Source Book in the Foundations of Mathematics}, Volume 2,
 (editor: W.B.\ Ewald),
Oxford University Press, Oxford, 1996, pp.~1115--1134;
\ISBN 0-19-850536-1.

\biblabel Hume (1748)
David Hume,
\book{An Enquiry Concerning Human Understanding};
\URL Project Gutenberg<https://archive.org/download/rmcg0003/Hume-Enquiry-a1.pdf>.

\biblabel Jackendoff (2011)
Ray Jackendoff,
\article{What Is the Human Language Faculty? Two Views};
in \periodical{Language},
vol.\ 87, no.\ 3, pp.\ 586--624, September 2011,
\DOI{10.1353/lan.2011.0063}.

\biblabel Kant (1783)
Immanuel Kant,
\book{Prolegomena to Any Future Metaphysics
 That Will Be Able to Present Itself as a Science};
translated by Paul Carus (1902),
\URL Project Gutenberg<https://archive.org/download/rmcg0004/Kant-Prolegomena-a1.pdf>.

\biblabel Kenneally (2007)
Christine Kenneally,
\book{The First Word: The Search for the Origins of Language};
Penguin Books, New York, 2008,
\ISBN 978-0-14-311374-4.

\biblabel Kim (2011)
Jaegwon Kim,
\book{Philosophy of Mind}, Third Edition;
Westview Press, Boulder, Colorado,
\ISBN 978-0-8133-4458-4,
e{\sc isbn}: 978-0-8133-4520-8.

\biblabel Kleene (1935)
Stephen Kleene,
\article{General Recursive Functions of Natural Numbers};
in \periodical{Mathematische Annalen},
vol.\ 112, no.\ 1, pp.\ 727--742, December 1936,\\
\DOI{10.1007/BF01565439}.
Presented to the American Mathematical Society,
September 1935.

\biblabel Kleene (1936)
Stephen Kleene,
\article{$\lambda$-Definability and Recursiveness};
in \periodical{Duke Mathematical Journal},
vol.\ 2, pp.\ 340--353, 1936,
\DOI{10.1215/s0012-7094-36-00227-2}.

\biblabel Kleene (1943)
Stephen Kleene,
\article{Recursive Predicates and Quantifiers};
in \periodical{Transactions of the American Mathematical Society},
vol.\ 53, no.\ 1, pp.\ 41--73, January 1943,
\DOI{10.2307/1990131}.

\biblabel Kleene (1952)
Stephen Kleene,
\book{Introduction to Meta-Mathematics};
Ishi Press, New York, 2009,
\ISBN 978-0-923891-57-2.
Reprint of the same title by
North-Holland, Amsterdam, 1952.

\biblabel Krifka (2011)
Manfred Krifka,
\article{Questions},
\DOI{10.1515/9783110255072.1742};\\
in \book{Semantics:
 An International Handbook of Natural Language Meaning},
 Volume 2, HSK 33.2,
 K.\ von Heusinger, C.\ Maienborn, P.\ Portner (editors),
pp.\ 1742--1785,
De Gruyter Mouton, Berlin/Boston, 2011,
\ISBN 978-3-11-018523-2.

\biblabel Kuhn (1970)
Thomas S.\ Kuhn,
\book{The Structure of Scientific Revolutions},
Second Edition, Enlarged;
The University of Chicago Press, Chicago, 1970,
\ISBN 978-0-226-45804-5.

\biblabel Lakatos (1976)
Imre Lakatos,
\book{Proofs and Refutations: The Logic of Mathematical Discovery},
edited by John Worrall and Elie Zahar;
Cambridge University Press, Cambridge UK, 1976,
\ISBN 0-521-29038-4.

\biblabel Lettvin et al. (1959)
Jerome Y.\ Lettvin, Humberto R.\ Maturana,
Warren S.\ McCulloch, and Walter H.\ Pitts,
\article{What the Frog's Eye Tells the Frog's Brain};
in \periodical{Proceedings of the IRE},
vol.\ 47, no.\ 11, pp.\ 1940--1951, November 1959,\\
\DOI{10.1109/jrproc.1959.287207}.

\biblabel Marr (1982)
David Marr,
\book{Vision: A Computational Investigation into
 the Human Representation and Processing of Visual Information};
W.H.\ Freeman and Company, San Francisco, CA, 1982,
\ISBN 0-7167-1284-9.

\biblabel Maturana \& Varela (1973)
Humberto Maturana Romesín, and
Francisco J.\ Varela García,
\book{De Máquinas y Seres Vivos: Autopoiesis: La Organización de lo Vivo},
Quinta edición; Editorial Universitaria, Santiago de Chile, 1994,
\ISBN 956-11-1211-6.

\biblabel McCarthy (1960)
John McCarthy,
\article{Recursive Functions of Symbolic Expressions
and Their Computation by Machine, Part I};
in \periodical{Communications of the ACM},
vol.\ 3, no.\ 4, pp.\ 184--195, April 1960,
\DOI{10.1145/367177.367199}.

\biblabel McCarthy et al. (1962)
John McCarthy, Paul Abrahams, Daniel Edwards,
Timothy Hart, and Michael Levin,
 \book{LISP 1.5 Programmer's Manual};
The MIT Press, Cambridge MA, 1962,
\ISBN 978-0-262-13011-0.

\biblabel McCulloch \& Pitts (1943)
Warren McCulloch and Walter Pitts,
\article{A Logical Calculus of the Ideas Immanent in Nervous Activity};
in \periodical{Bulletin of Mathematical Biophysics},
vol.\ 5, no.\ 4, pp.\ 115--133, 1943,
\DOI{10.1007/BF02478259}.

\biblabel Miller (1956)
 George Miller,
\article{The Magical Number Seven, Plus or Minus Two:
Some Limits on Our Capacity for Processing Information};
in \periodical{Psychological Review},
vol.\ 63, No.\ 2, pp.\ 81--97, 1956,
\DOI{10.1037/h0043158}.

\biblabel Nelson (1987)
R.\ J.\ Nelson,
\article{Church's Thesis and Cognitive Science};
in \periodical{Notre Dame Journal of Formal Logic},
vol.\ 28, no.\ 4, pp.\ 581--614, October 1987,\\
\DOI{10.1305/ndjfl/1093637649}.

\biblabel Olszewsky (2005)
Adam Olszewski,
\article{Church's thesis as an empirical hypothesis};
in \periodical{Annales UMCS Informatica},
vol.\ AI 3, pp.\ 119--130, 2005,
\DOI{10.17951/ai.2005.3.1.119-130},
\URL<https://journals.umcs.pl/ai/article/view/3013>.

\biblabel Ouattara et al. (2009)
Karim Ouattara, Alban Lemasson, and Klaus Zuberbühler,\\
\article{Campbell's monkeys concatenate vocalizations
 into context-specific call sequences};
in \periodical{PNAS},
vol.\ 106, no.\ 51, pp.\ 22026--22031, December 22, 2009,\\
\DOI{10.1073/pnas.0908118106}.

\biblabel Pais (1982)
Abraham Pais,
\book{‘Subtle is the Lord \dots’:
 The Science and the Life of Albert Einstein};
Oxford University Press, Oxford, 1982,
\ISBN 0-19-285138-1.

\biblabel Patterson et al. (2006)
N.\ Patterson, D.J.\ Richter, S.\ Gnerre, E.S.\ Lander, and D.\ Reich,
\article{Genetic evidence for complex speciation of humans and chimpanzees};
in \periodical{Nature},
vol.\ 441, pp.\ 1103--1108, 2006,
\DOI{10.1038/nature04789}.

\biblabel Pavlov (1927)
Ivan Pavlov,
\book{Conditioned Reflexes:
 An Investigation of the Physiological Activity
 of the Cerebral Cortex},
translated by G.V.\ Anrep;
Oxford University Press, London, 1927.
Reprinted by Dover, Minneola, NY, 1960, 2003;
\ISBN 0-486-43093-6.

\biblabel Penrose (1989)
Roger Penrose, \negthinspace
\book{The Emperor's New Mind: \negthinspace
Concerning Computers, Minds, and the Laws of Physics};
Oxford University Press, Oxford, 1989,
\ISBN 0-19-851973-7.

\biblabel Pierce (1867)
Charles Sanders Peirce,
\article{On a New List of Categories};
in \periodical{Proceedings of the American Academy
of Arts and Sciences},
vol.\ 7, pp.\ 287--298, 1868,
\DOI{10.2307/20179567}.
Presented May 14, 1867.

\biblabel Pinker \& Jackendoff (2004)
Steven Pinker, and Ray Jackendoff,
\article{The Faculty of Language: What’s Special About It?};
in \periodical{Cognition},
vol.\ 95, pp.\ 201--236, 2005,
\DOI{10.1016/j.cognition.2004.08.004}.
Received 16 January 2004; accepted 31 August 2004.

\biblabel Post (1936)
Emil L.\ Post,
\article{Finite Combinatory Processes --- Formulation 1};
in \periodical{The Journal of Symbolic Logic},
Volume\ 1, Number\ 3, pp.~103--105, September 1936,
\DOI{10.2307/2269031}.
Received October 7, 1936.

\biblabel Post (1944)
Emil L.\ Post,
\article{Recursively Enumerable Sets of Positive Integers
  and their Decision Problems};
in \periodical{Bulletin of the American Mathematical Society},
vol.\ 50, no.~5, pp.\ 284--316, 1944,
\DOI{10.1090/s0002-9904-1944-08111-1}.

\biblabel Progovac (2015)
Ljiljana Progovac,
\book{Evolutionary Syntax};
Oxford University Press, Oxford, 2015,
\ISBN 978-0-19-873655-4.

\biblabel Progovac (2016)
Ljiljana Progovac,
\article{A Gradualist
Scenario for Language Evolution:
Precise Linguistic Reconstruction of
Early Human (and Neandertal) Grammars};
in \periodical{Frontiers in Psychology},
vol.\ 7, art.\ 1714, 2016,
\DOI{10.3389/fpsyg.2016.01714}.

\biblabel Putnam (1975)
Hilary Putnam,
\article{The Meaning of ‘Meaning’$\,$};
in \periodical{Minnesota Studies in the Philosophy of Science},
Volume 7, Number 3, pages 131--193, 1975.
Reprinted in
\book{Philosophical Papers, Volume 2: Mind, Language and Reality},
edited by Hilary Putnam,
Cambridge University Press, pp.\ 215--271, 1975,
\ISBN 978-0-521-29551-2, with
\DOI{10.1017/CBO9780511625251.014}.

\biblabel Putnam (1988)
Hilary Putnam,
\book{Representation and Reality};
The MIT Press, Cambridge, MA, 1988,
\ISBN 978-0-262-66074-7.

\biblabel Rosen (1985)
Robert Rosen,
\book{Anticipatory Systems:
 Philosophical, Mathematical, and Methodological Foundations},
 IFSR International Series on
 Systems Science and Engineering, Volume 1;
 Pergamon Press, Oxford, 1985,
\ISBN 0-08-031158-X.

\biblabel Rosser (1936)
J.\ Barkley Rosser,
\article{Extensions of Some Theorems of Gödel and Church};
in \periodical{Journal of Symbolic Logic},
vol.\ 1, pp.\ 87--91, 1936,
\DOI{10.2307/2269028}.

\biblabel Schlosshauer et al. (2013)
Maximilian Schlosshauer, Johannes Kofler, and Anton Zeilinger,
\article{A Snapshot of Foundational Attitudes Toward Quantum Mechanics};
in \periodical{Studies in History and Philosophy of Science Part B:
 Studies in History and Philosophy of Modern Physics},
vol.\ 44, no.\ 3, pp.\ 222--230, August 2013,
DOI{10.1016/j.shpsb.2013.04.004},
URL<https://arxiv.org/abs/1301.1069>.

\biblabel Searle (1980)
John Searle,
\article{Minds, Brains and Programs};
in \periodical{Behavioral and Brain Sciences},
vol.\ 3, no.\ 3, pp.\ 417--457, September 1980,
\DOI{10.1017/S0140525X00005756}.

\biblabel Searle (1992)
John R.\ Searle,
\book{The Rediscovery of the Mind};
The MIT Press, Cambridge, MA, 1992,
\ISBN 978-0-262-69154-3.

\biblabel Seyfarth et al. (1980)
Robert M.\ Seyfarth, Dorothy L.\ Cheney, and Peter Marler,
\article{Monkey Responses to Three Different Alarm Calls:
  Evidence of Predator Classification and Semantic Communication};
in \periodical{Science},
vol.\ 210, no.\ 4471, pp.\ 801--803, November 1980,
\DOI{10.1126/science.7433999}.

\biblabel Shagrir \& Pitowsky (2003)
Oron Shagrir and Itamar Pitowsky,
\article{Physical Hypercomputation and the Church-Turing Thesis};
in \periodical{Minds and Machines},
vol.\ 13, pp.\ 87--101, 2003,
\DOI{10.1023/A:1021365222692}.

\biblabel Shannon (1948)
C.\ E.\ Shannon,
\article{A Mathematical Theory of Communication};
in \periodical{Bell System Technical Journal},
vol.\ 27, no.\ 3 \& no.\ 4, pp.\ 379--423 \& 623--656,
\DOI{10.1002/j.1538-7305.1948.tb01338.x} \&
\DOI{10.1002/j.1538-7305.1948.tb00917.x}.

\biblabel Shieber (1986)
Stuart Shieber,
\book{An Introduction to Unification-Based Approaches to Grammar};
Microtome Publishing, Brookline, Massachusetts, 2003.
Reissue of the same title by
CSLI Publications, Stanford, California, 1986.\\
{\sc url:} \URL<http://nrs.harvard.edu/urn-3:HUL.InstRepos:11576719>.

\biblabel Simon \& Newell (1971)
Herbert Simon and Allen Newell,
\article{Human Problem Solving: The State of the Theory in 1970};
in \periodical{American Psychologist},
 vol.\ 26, no.\ 2, pp.\ 145--159, February 1971,
\DOI{10.1037/h0030806}.

\biblabel Stabler (2014)
Edward Stabler,
\article{Recursion in Grammar and Performance},\\
\DOI{10.1007/978-3-319-05086-7_8};
in \book{Recursion: Complexity in Cognition},
Volume 43 of the series ‘Studies in Theoretical Psycholinguistics’,
Tom Roeper, Margaret Speas (editors),
pp.\ 159--177,
Springer, Cham, Switzerland, 2014,
\ISBN 978-3-319-05085-0.

\biblabel Tomasello (2008)
Michael Tomasello,
\book{Origins of Human Communication};
The MIT Press, Cambridge, Massachusetts, 2008,
\ISBN{978-0-262-51520-7}.

\biblabel Tomasello (2009)
Michael Tomasello.
\article{Universal Grammar Is Dead};
in \periodical{Behavioral and Brain Sciences},
vol.\ 32, no.\ 5, pp.\ 470--471, 2009,
\DOI{10.1017/S0140525X09990744}.

\biblabel Turing (1936)
A.\ M.\ Turing,
\article{On Computable Numbers,
 with an Application to the Entscheidungsproblem};
in \periodical{Proceedings of the London Mathematical Society},
vol.\ s2-42, no.\ 1, pp.\ 230--265, 1937,
\DOI{10.1112/plms/s2-42.1.230}.
Received 28 May, 1936. Read 12 November, 1936.

\biblabel Turing (1937)
Alan Turing,
\article{Computability and $\lambda$-Definability}; in
\periodical{The Journal of Symbolic Logic},
vol.\ 2, no.\ 4, pp.\ 153--163, December 1937,
\DOI{10.2307/2268280}.

\biblabel Turing (1938)
Alan Turing,
\article{Systems of Logic Based on Ordinals};
in \periodical{Proceedings of the London Mathematical Society},
vol.\ s2-45, no.\ 1, pp.\ 161--228, 1939,\\
\DOI{10.1112/plms/s2-45.1.161}.
Received 31 May, 1938. Read 16 June, 1938.

\biblabel Vygotsky (1934)
Lev Vygotsky,
\book{Thought and Language},
newly revised and edited by Alex Kozulin;
The MIT Press, Cambridge MA, 1986,
\ISBN 978-0-262-72010-6.

\biblabel Watumull et al. (2014)
Jeffrey Watumull, Marc Hauser, Ian Roberts \& Norbert Hornstein,
\article{On Recursion}; in
\periodical{Frontiers in Psychology},
vol.\ 4, article 1017, pp.\ 1--7, 2014,
\DOI{10.3389/fpsyg.2013.01017}.

\biblabel Wilson (2008)
Edward O.\ Wilson,
\article{One Giant Leap: How Insects Achieved Altruism
 and Colonial Life};
in \periodical{BioScience},
vol.\ 58, no.\ 1, pp.\ 17--25, January 2008,
\DOI{10.1641/B580106}.

\biblabel Wittgenstein (1922)
Ludwig Wittgenstein,
\book{Tractatus logico-philosophicus};
Routledge \& Kegan Paul, London, 1922.

\biblabel Zermelo (1908)
Ernst Zermelo,
\articlede{Untersuchungen über die Grundlagen der Mengenlehre I},
\DOI{10.1007/978-3-540-79384-7_6};
in \book{Ernst Zermelo Collected Works} Volume I,
H.-D.\ Ebbinghaus, A.\ Kanamori (editors),
pp.\ 160--229,
Springer-Verlag, Berlin Heidelberg, 2010,
\ISBN 978-3-540-79383-0.

 \let\item=\bibitem \let\xitem=\xbibitem
 \everypar{}\smallskip \parskip=0pt \frenchspacing}
\def\endreferences{\par\endgroup}

\def\bibitem #1 (#2){\chreference #1 (#2):\enskip
 {\stringaccents\def\&{\string\&}\xdef\1{#1}}\csname \1#2\endcsname}
\def\xbibitem #1 (#2) = #3 (#4){\chreference #1 (#2):\enskip
 {\stringaccents\def\&{\string\&}\xdef\1{#3}}\csname \1#4\endcsname}

\def\book#1{{\em#1}}
\def\article#1{“#1”}
\def\articlede#1{„#1“}
\def\periodical#1{{\em#1}}
\def\ISBN{{\sc isbn: }}

\catcode`\_12 \def\@nder{_} \catcode`\_13
\def\DOI{{\sc doi: }\begingroup\catcode`\_=13\relax\d@i}
\def\d@i#1{\edef\1{#1}\relax\let_\@nder\relax\edef\2{\1<https://doi.org/#1>}%
 \expandafter\URL\2\endgroup}
\catcode`\_=8



\newdimen\tochsize \tochsize=\hsize \advance\tochsize-5\baselineskip
\newdimen\tocindent \tocindent=2\baselineskip 
\def\tocbox{\hbox to\tocindent}

\def\leaderfill{\leaders\hbox to\baselineskip{\hss.\hss}\hfill}
\def\unfakeablepar{\unskip\nobreak
 \leaders\hbox to\baselineskip{\hss.\hss}\hskip\parfillskip
 \vadjust{}{\parfillskip=-\baselineskip\endgraf}}

\def\toclineX#1#2#3#4#5{\count@=#1\def\0{#4}%
 \ifx\0\star \csname#2\endcsname \else
 \ifx\0\lq \relax \else
 \ifx\0\rq \lrelax \else
 \ifx\0\empty \let\0=\ignorespaces \else
 \ifx\0\lbrack \let\0=\ignorespaces \else
 \ifx\0\rbrack \let\0=\ignorespaces \else
 \def\0{\tocbox{#4\hfil}}\fi\fi\fi
 \setbox0=\hbox{\kern\count@\tocindent\0#2}%
 \ifdim\wd0>\tochsize \advance\count@1
  \line{\vbox{\hsize=\tochsize \rightskip=0pt plus6em
   \hangindent\count@\tocindent \noindent\strut
    \kern#1\tocindent\0#2\strut\unfakeablepar}\leaderfill
  {\xiibf\goto{#5}{#3}}}%
 \else
  \line{\strut\kern#1\tocindent\0#2\leaderfill {\xiibf\goto{#5}{#3}}}%
 \fi
 \fi\fi\fi}

\def\toclineC#1#2#3#4#5{\def\0{#4}%
 \ifx\0\star \csname#2\endcsname \else
 \ifx\0\lq \relax \else
 \ifx\0\rq \lrelax \else
 \ifx\0\empty \let\0=\ignorespaces \else
 \ifx\0\lbrack \let\0=\ignorespaces \else
 \ifx\0\rbrack \let\0=\ignorespaces \fi\fi\fi
 \begincenter\goto{\0 #2}{#3}\endcenter
 \fi\fi\fi}

\def\toclineR#1#2#3#4#5{\ifcase#1
  \bigbreak{\fonttwo\toclineC{#1}{#2}{#3}{#4}{#5}}\medskip\or
  \toclineX{0}{#2}{#3}{#4}{#5}\or
  \def\0{#4}\ifx\0\empty\else \toclineX{1}{#2}{#3}{#4}{#5}\fi
  \else\relax\fi}

\let\tocline=\toclineR

\newread\tocfile
\def\maketoc{\openin\tocfile=auxiliar.toc
 \ifeof\tocfile
  \closein\tocfile
  \message{No toc file!}%
 \else
  \closein\tocfile
  \dest\let\secid\lq\toc1{Contents}%
  \tochsize=\hsize \advance\tochsize-5\baselineskip
  \input auxiliar.toc
 \fi}


\def\strutdepth{\dp\strutbox}
\def\marginalstar{\strut\vadjust{\kern-\strutdepth\specialstar}}
\let\marginalsymbol=*
\def\specialstar{\vtop to \strutdepth{
 \baselineskip\strutdepth
 \vss\llap{\rm\truecolor{Red}\marginalsymbol\endcolor\quad}\null}}

\def\new/{\truecolor{Red}}
\def\wen/{\endcolor\ifhmode\marginalstar\fi}

\def\uncatcodespecials{\def\do##1{\catcode`##1=12 }\dospecials}
\def\del/{\begingroup\uncatcodespecials
 \ifvmode \let\next=\DELv \else \let\next=\DELh \fi\next}
{\catcode`\|=0 \catcode`\\=12
 |long|gdef|DELv#1\led/{|endgroup}
 |long|gdef|DELh#1\led/{|marginalstar|endgroup|ignorespaces}}
\def\led/{\errmessage{Error! Unnested led.}}



\def\uncatcodespecials{\def\do##1{\catcode`##1=12 }\dospecials}

\def\verb{\begingroup\setupverbatim\d@verbatim}
\def\setupverbatim{\tt\parskip0pt plus0.1pt minus0.1pt
 \everypar={}\def\par{\leavevmode\endgraf}\catcode`\`=13
 \uccode`\~=`\ \uppercase{\let~=\ }\uccode`\~=`\`
 \uppercase{\def~{\relax\lq}}\uccode`\~=0
 \obeylines\uncatcodespecials\obeyspaces \verbatimoptions}
\def\d@verbatim#1{\def\next##1#1{##1\endgroup}\next}
\def\verbatimoptions{}
{\catcode`\`=13 \gdef`{\lq}} 


\def\newline{\ifvmode\null\else\null\hfil\break\fi}
\let\\=\newline

\def\beginpoints{\begingroup\parskip=0pt\parindent=20pt\everypar{}}
\def\point{\futurelet\nexttoken\pointxx}
\def\pointxx{\if [\nexttoken \let\next=\namedpoint \else \let\next=\defaultpoint \fi\next}
\def\namedpoint[#1]{\setbox0=\hbox{#1}\ifdim\wd0<\parindent \item{#1}\else \par\hang\noindent{#1}\fi}
\def\defaultpoint{\item{$\circ$}}
\def\endpoints{\par\noindent\endgroup}

\def\begincitation{\smallskip\begingroup\everypar={}\sl
 \advance\leftskip20pt\advance\rightskip20pt}
\def\endcitation{\smallskip\noindent\endgroup}

\def\em{\ifdim \fontdimen1\font=0pt \aftergroup\smartitc@r \fi \it}
\def\qt{\ifdim \fontdimen1\font=0pt \aftergroup\smartitc@r \fi \sl}
\def\smartitc@r{\ifhmode \expandafter\itpuncl@ok \fi}
\def\itpuncl@ok{\begingroup\futurelet\ITCt@mpa\itcort@st}
\def\itcort@st{\def\ITCt@mpb{\ITCt@mpa}%
 \ifcat\noexpand\ITCt@mpa,\setbox0=\hbox{\ITCt@mpb}%
  \ifdim\ht0<0.3ex \let\itc@rdo=\endgroup \fi\fi \itc@rdo}
\def\itc@rdo{\skip0=\lastskip \ifdim\skip0=0pt \/\else
 \unskip \/\hskip\skip0 \fi \endgroup}

\def\definition#1{{\em #1}}
\def\latin#1{{\em #1}}

\def\URL{\leavevmode\begingroup\catcode`\#=12\catcode`\_=12\relax\@RL}
\def\@RL#1<#2>{\def\1{#1}\ifx\1\empty\def\2{#2}\else\def\2{#1}\fi
\ifcase\pdfoutput
 \ifx\dvips\undefined
  {\color{Red}\2\endcolor}\else
  {\color{Red}\2\endcolor}%
\fi
\else
 \pdfstartlink attr{/Border [0 0 0]}
  user{/Subtype /Link /A << /Type /Action
  /S /URI /URI (#2) >>}{\color{Red}\2\endcolor}\pdfendlink
\fi\endgroup\relax}

\def\MTfigures{\ifx\MT\undefined 

%
%
%

\message{V1.5 by RMCG 20221206}  

\chardef\MToldatcatcode=\catcode`\@\catcode`\@=11

\newif\ifMTf@le 
\newif\ifMTmf 
\newwrite\MToutf@le
\newread\MTinf@le
\newbox\MTbox
\newbox\MTb@x
\newcount\MTn@
\newdimen\MTxpos@
\newdimen\MTypos@

\openin\MTinf@le=auxiliar.mf 
\ifeof\MTinf@le \MTf@lefalse \else \MTf@letrue \fi
\closein\MTinf@le

\def\MTendmark{}
{\obeylines\gdef\MTign@re#1
 {\def\next{#1}\ifx\next\MTendmark \let\next\endgroup \else %
  \let\next\MTign@re\fi \next}}
{\obeylines\gdef\MTc@py#1
 {\def\next{#1}\ifx\next\MTendmark \let\next\endgroup \else %
  \immediate\write\MToutf@le{\next}\let\next\MTc@py\fi \next}}

{\obeylines\gdef\MTign@reline#1
 {\endgroup}}
{\obeylines\gdef\MTc@pyline#1
 {\immediate\write\MToutf@le{#1}\endgroup}}

\def\MTsetupc@py{\def\do##1{\catcode`##1=12 }\dospecials
 \catcode`\\=0 \let\\=\MTbackslash \obeyspaces\obeylines}

\def\MTmf@mp{\ifeof\MTinf@le
 \errhelp={It seems to me that file auxiliar.mf has not been processed,
           or that file auxiliar.log was deleted.}%
 \errmessage{File auxiliar.log not found}%
 \else
  \read\MTinf@le to \next 
  \def\MT@mf{METAFONT}%
  \def\MT@##1 ##2 ##3,##4\MT@@{##3}
  \edef\MTmeta{\expandafter\MT@\next\MT@@}%
  \message{(Drawings made by \MTmeta)}%
  \ifx\MTmeta\MT@mf \MTmftrue \else \MTmffalse \fi
 \fi}

\ifx\pdfoutput\undefined \csname newcount\endcsname\pdfoutput \fi

\def\MTl@gl@b{\MTloglabel }
\def\MTgetl@g{\ifeof\MTinf@le \let\next\relax
  \errhelp={I was expecting to read a label location.}%
  \errmessage{Unexpected end of auxiliar.log}%
 \else \read\MTinf@le to \next
 \ifx\next\MTl@gl@b \let\next\relax
 {\catcode`\>=9
  \global\read\MTinf@le to \MTxpos@text
  \global\read\MTinf@le to \MTypos@text }%
 \else \let\next\MTgetl@g \fi\fi \next}

\escapechar=-1 \edef\MTbackslash{\string\\}\escapechar=`\\
\def\MTslashing{\begingroup \escapechar=-1 \edef\\{\string\\}%
 \edef\{{\string\{}\edef\}{\string\}}\edef\#{\string\#}%
 \edef\${\string\$}\edef\^{\string\^}\edef\_{\string\_}%
 \edef\&{\string\&}\edef\~{\string\~}\edef\%{\string\%}%
 \escapechar=`\\}
\let\MTendslashing=\endgroup

\def\MTf@rst#1#2#3/{#1}\def\MTsec@nd#1#2#3/{#2}

\ifMTf@le \message{Second TeX pass} 
 \openin\MTinf@le=auxiliar.log
 \MTmf@mp
 \ifMTmf 
  \font\MTfont=auxiliar \def\MTchar{\MTfont\char\MTn@}%
 \else   
  \ifcase\pdfoutput
   \def\MTchar{\includegraphics{auxiliar.\number\MTn@}}%
  \else
   \input mtmp2pdf
   \def\MTchar{\MPtoPDF{auxiliar.\number\MTn@}}%
  \fi
 \fi
 \def\MTcode{\begingroup\MTsetupc@py\MTign@re{ }}
 \def\MT:{\begingroup\MTsetupc@py\MTign@reline}
 \def\MTline#1{}
 \def\MTbeginchar(#1,#2,#3);{\setbox\MTb@x=\hbox{\MTchar}%
  \wd\MTb@x=#1\ht\MTb@x=#2\dp\MTb@x=#3%
  \global\setbox\MTbox=\vtop{\box\MTb@x}}
 \def\MTlabel#1(#2)#3"#4";{\setbox\MTb@x\hbox{#4}\MTgetl@g
  \MTxpos@=\MTxpos@text pt \MTypos@=\MTypos@text pt \dimen@=\dp\MTbox
  \global\setbox\MTbox=\vtop{\unvbox\MTbox\nointerlineskip
   \def\1{\MTf@rst#3cc/}\def\2{\MTsec@nd#3cc/}%
   \vbox to 0pt{\advance\dimen@\MTypos@ \kern-\dimen@
    \if b\1\kern-\ht\MTb@x \kern-\dp\MTb@x \else
     \if c\1\kern-0.5\ht\MTb@x \kern-0.5\dp\MTb@x \else
      \if g\1\kern-\ht\MTb@x \fi\fi\fi
    \hbox to 0pt{\kern\MTxpos@
     \if r\2\kern-\wd\MTb@x \else \if c\2\kern-0.5\wd\MTb@x \fi\fi
     \box\MTb@x\hss}\vss}}}
 \def\MPlabel#1(#2)#3"#4";{\relax}
 \def\MTendchar;{\global\advance\MTn@1 }
\else \message{First TeX pass} 
 \let\MTfont=\nullfont
 \let\MTchar=\relax
 \tracinglostchars=0
 \immediate\openout\MToutf@le=auxiliar.mf
 \def\MTcode{\begingroup\MTsetupc@py\MTc@py{ }}
 \def\MT:{\begingroup\MTsetupc@py\MTc@pyline}
 \def\MTline#1{\MTslashing\immediate\write\MToutf@le{#1}\MTendslashing}
 \def\MTbeginchar(#1,#2,#3);{%
  \MTline{beginchar(\number\MTn@,#1\#,#2\#,#3\#); \% line \the\inputlineno}%
  \global\setbox\MTbox=\vtop{}\wd\MTbox=#1\ht\MTbox=#2\dp\MTbox=#3}
 \def\MTlabel#1(#2)#3"#4";{{\setbox0=\hbox{#4}%
  \def\1{\MTf@rst#3cc/}\def\2{\MTsec@nd#3cc/}%
  \MTline{ MTlabel(#2)("\1","\2","#1",\the\wd0,\the\ht0,\the\dp0);}}}
 \def\MPlabel#1(#2)#3"#4";{{\def\1{#1}\ifx\1\empty\else\def\1{x}\fi
  \def\@tl@{.lrt}\def\@tc@{.bot}\def\@t@{.bot}\def\@tr@{.llft}\let\@c@=\empty
  \def\@cl@{.rt}\def\@cr@{.lft}\let\@cc@=\empty
  \def\@bl@{.urt}\def\@bc@{.top}\def\@b@{.top}\def\@br@{.ulft}\let\@@=\empty
  \def\3{\csname @#3@\endcsname}%
  \MTline{ \1label\3(btex #4 etex, z#2);}}}
 \def\MTendchar;{\MTline{endchar;}\MTline{}\global\advance\MTn@1 }
\fi

\MT:
\MT:
\MT: mag:=\number\mag/1000;
\MT:
\MT: truept = pt/mag; truept# = pt#/mag;
\MT: truepc = pc/mag; truepc# = pc#/mag;
\MT: truein = in/mag; truein# = in#/mag;
\MT: truebp = bp/mag; truebp# = bp#/mag;
\MT: truecm = cm/mag; truecm# = cm#/mag;
\MT: truemm = mm/mag; truemm# = mm#/mag;
\MT: truedd = dd/mag; truedd# = dd#/mag;
\MT: truecc = cc/mag; truecc# = cc#/mag;
\MT:
\MT:if known prologues: 
\MT:
\MT: prologues := 3;
\MT:
\MT: mm#=2.84528;      pt#=1;        dd#=1.07001;      bp#=1.00375;
\MT: cm#=28.45276;     pc#=12;       cc#=12.84010;     in#=72.27;
\MT:
\MT: bboxmargin := 0pt; labeloffset := 0bp;
\MT:
\MT: string extra_setup, extra_beginchar, extra_endchar;
\MT: extra_setup = extra_beginchar = extra_endchar = "";
\MT:
\MT: def mode_setup =
\MT:  proofing:=0;                   
\MT:  fontmaking:=0;                 
\MT:  tracingtitles:=0;              
\MT:  scantokens extra_setup;
\MT: enddef;
\MT:
\MT: def beginchar(expr c,w_sharp,h_sharp,d_sharp) =
\MT:  begingroup
\MT:   charcode:=c;
\MT:   w:=w_sharp*pt; h:=h_sharp*pt; d:=d_sharp*pt;
\MT:   clearxy; clearit; clearpen;
\MT:   pickup defaultpen;
\MT:   drawoptions();
\MT:   scantokens extra_beginchar;
\MT: enddef;
\MT:
\MT: def endchar =
\MT:   scantokens extra_endchar;
\MT:   shipit;
\MT:  endgroup
\MT: enddef;
\MT:
\MT: def xlabel text t =
\MT:  erase fill bbox thelabel t;
\MT:  draw thelabel t;
\MT: enddef;
\MT:
\MT: vardef xlabels@#(text t) =
\MT:  forsuffixes $=t:
\MT:    xlabel@#(str$,z$); endfor
\MT:  enddef;
\MT:
\MT:else: 
\MT: mode:=localfont;
\MT:fi
\MT:
\MT:mode_setup;
\MT:
\MT:def MTlabel(suffix i)(expr v,h,add,wd,ht,dp) =
\MT: x.i.l = x.i if h="c": - wd/2 elseif h="r": - wd fi;
\MT: y.i.t = y.i if v="c": + (ht+dp)/2 elseif v="b": + ht+dp elseif v="g": + ht fi;
\MT: x.i.r = x.i.l + wd; y.i.b = y.i.t - ht - dp;
\MT: if not (add=""): erase fill (x.i.l,y.i.t) --
\MT:  (x.i.l,y.i.b) -- (x.i.r,y.i.b) -- (x.i.r,y.i.t) -- cycle; fi
\MT: message"\\MTloglabel"; show x\\i/pt; show y\\i/pt; message"";
\MT:enddef;
\MT:

\let\texbye=\bye
\outer\def\bye{\ifMTf@le \closein\MTinf@le \else
 \MTline{end.}\immediate\closeout\MToutf@le \fi
 \par\vfill\supereject\end}

\catcode`\@=\MToldatcatcode
 \fi
 \let\MTbeginfig=\MTbeginchar \let\MTendfig=\MTendchar 
 \MTline{if known prologues:} 
 \MTline{ verbatimtex}
 \MTline{ \\input fonts \\xiifonts}
 \MTline{ \\catcode`\\@=11} 
 \MTline{ etex}
 \MTline{fi}
 \MTline{}}

\catcode`\@=12

\darkcolors              
\MTfigures

\def\version{20230321}  

\def\thesis(#1)#2{$$\hbox{\it #2}\leqno\diamond\ \hbox{Thesis #1}$$}


\title{Proof of Church's Thesis}
\author{Ramón Casares}
\contact{{\sc orcid:}
\URL 0000-0003-4973-3128<https://orcid.org/0000-0003-4973-3128>}
\subject{Mathematics}
\keywords{Church's thesis,  Post (1936) program,
  Post's law, Turing completeness.}

\beginnote
This is \DOI{10.6084/m9.figshare.4955501}.v4,
version {\tt\version}.\\
Alternative \DOI{10.48550/arXiv.1209.5036}.\\
\copyright\ 2012--2023 Ramón Casares; licensed as {\tt cc-by}.\\
Any comments on it to
{\tt\URL papa@ramoncasares.com<mailto:papa@ramoncasares.com>}
are welcome.
\endnote

\maketitle

\beginabstract
We prove that
if our calculating capability is
that of a universal Turing machine with a finite tape,
then Church's thesis is true.
This way we accomplish \cite Post (1936) program.
\endabstract

\section{Church's Thesis}

Church's thesis, also known as Church-Turing thesis,
says, see \cite Gandy (1980):
\thesis(1){What is effectively calculable is computable.}

\label{Thesis1}{Euler Diagram of Church's Thesis}
\MTbeginfig(150bp, 70bp, 0bp);  
\MT: d := 3bp;
\MT: pickup pencircle scaled 1 bp;
\MT: z0 = (w/2,2h/3);
\MTlabel(0)"Effectively calculable";
\MT: draw (x0l-d,y0b-d) ... (w/2,y0b-3d) ... (x0r+d,y0b-d) ...
\MT:      (x0r+d,y0t+d) ... (w/2,y0t+4d) ... (x0l-d,y0t+d) ... cycle;
\MT: z1 = (w/2, h/3 - 2d);
\MTlabel(1)"Computable";
\MT: draw (x0l-2d,y1b-d) ... (w/2, y1b-3d) ... (x0r+2d,y1b-d) ...
\MT:      (x0r+2d,y0t+2d) ... (w/2, y0t+6d) ... (x0l-2d,y0t+2d) ... cycle;
\MT: 
\MT: 
\MTendfig;
\continuepar$$\box\MTbox$$

What is \definition{computable} is anything that
any Turing machine can compute,
where the Turing machine was defined by \cite Turing (1936).
There are other definitions of computable,
using for example Church's $\lambda$-calculus,
but all of them are mathematically equivalent.
In any case,
the term computable is defined with mathematical rigor.

However, because what can be effectively calculated
was considered a vague notion,
it was assumed that Church's thesis could not be formally proved.
In what follows, we will make some definitions and assumptions
in order to overcome the ambiguity.
Then we will deduce Church's thesis from our
calculating limitations,
which we will postulate are those of a
universal Turing machine with a finite tape.
In doing so, we are realizing the old program of \cite Post (1936),
for whom Church's thesis is neither a definition nor an axiom,
but a {\it natural law\/} stating “the limitations of the
mathematicizing power of [our species {\it Homo sapiens\/}]”.

\section{Syntax Engine}

\definition{Syntax}, as opposed to semantics, is concerned with
transformations of strings of symbols,
irrespective of the symbols meanings,
but according to a precise and finite set of well-defined rules.
We call this set of rules \definition{algorithm}.

The previous definition of syntax
generalizes the linguistic one by \cite Chomsky (1957):
“Syntax is the study of the principles and processes
by which sentences are constructed in particular languages”
(page 11).
A sentence is a string of words, and
a word is a particular case of symbol.

We will assume here that persons, 
that is, the members of our own species \latin{Homo sapiens},
have a syntactic capability.
If a person speaks a particular language, then
she has a syntactic capability
according to the linguistic definition.
Also, most of mathematics is pure syntax
according to the general definition, see \cite Hilbert (1922),
and mathematics are produced and consumed by persons. 

We will call whatever that implements a syntactic capability
a \definition{syntax engine}.
So, if persons have a syntactic capability,
then each person has a syntax engine.
In the case of a general-purpose computer,
the central processing unit ({\sc cpu}) is its syntax engine.

\section{Finite Turing Machine}

Thus, the Turing machine can be seen as
a mathematical model for syntax engines, see \cite Chomsky (1959).
The only part of a Turing machine that cannot be
built physically is its infinite tape.
So we will define
a \definition{finite Turing machine}
as a Turing machine with a finite tape
instead of the infinite tape.
The tape is just read and write memory.

For each finite Turing machine
there is one \definition{corresponding Turing machine},
which is the Turing machine that is identical to the
finite one, except for the tape.
Defining \definition{processor} as the whole Turing machine
except its tape, then any finite Turing machine
and its corresponding Turing machine have the same processor.

If two or more finite Turing machines
have the same corresponding Turing machine,
then we say that they are \definition{equivalent}.
Equivalent finite Turing machines only difference
is the length of their tapes;
they have all the same processor.
And now, the corresponding Turing machine
of each equivalence class of finite Turing machines
can be seen as its ideal representative.

The processor has some finite memory itself to store
what \cite Turing (1936) called the “$m$-configurations”.
From the processor point of view,
this $m$-configurations memory is internal memory
while the tape is external memory.
Thus, basically, a processor is a finite-state automaton,
a Turing machine is a finite-state automaton
attached to an infinite external memory,
and a finite Turing machine is a finite-state automaton
attached to a finite external memory.

Because a universal Turing machine is a Turing machine,
see \cite Turing (1936),
we can define finite universal Turing machines the same way.
A \definition{finite universal Turing machine} is
a universal Turing machine, but with a finite tape
instead of the infinite tape.
The corresponding Turing machine of a
finite universal Turing machine is a universal Turing machine;
both have the same processor.
Nevertheless, as noted below in \refsc{Effective Computation},
finite universal Turing machines are not universal,
because they are finite.

\vfil\break

We will say that
a computing device is \definition{Turing complete}
if and only if
its corresponding Turing machine is a universal Turing machine.
In other words, ‘Turing complete device’
is synonymous with ‘finite universal Turing machine’.
However, although they are formally equivalent,
we would rather use the former to qualify real devices
and the latter to refer to the mathematical concept.

\section{Effective Computation}

We will call
any computation done by a finite Turing machine
in a finite time
an \definition{effective computation}.
In principle, effective computations are physically achievable,
since they do not require unbounded resources.

Finite Turing machines will fail on those computations
that require more tape than they have available.
Also, in practical terms, and particularly if the tape is long,
time available could be exhausted
before reaching a tape end or a {\sc halt} instruction,
and then the computation would have to be aborted, failing.
These computations that fail in the finite Turing machine
because of a lack of tape or a lack of time
would continue in its corresponding Turing machine,
and they will eventually succeed, or not.
Summarizing: computing is more successful than effectively
computing.

But, how much successful?
A computation is \definition{successful}
when the Turing machine has reached
a {\sc halt} instruction in a finite time.
And, whenever a computation has reached
a {\sc halt} instruction in a finite time,
the Turing machine has only had time to inspect a
finite number of tape cells. This means that
any successful computation could be done
by some finite Turing machine in a finite time.
So the answer is: not too much.

In addition,
each finite Turing machine computation that {\sc halt}s
will be run identically by its corresponding Turing machine,
because no limitations of tape or time were found,
and there are not any other differences between the two machines.
So any computation done by a finite Turing machine
will be identically computed by its corresponding Turing machine.
Summarizing: an effective computation is a computation.

The \definition{effective computation proposition}
is longer, and it is also mathematically precise:
{\it save for time or tape limitations,
every finite Turing machine computes exactly as
its corresponding Turing machine.}

A finite universal Turing machine is not
a universal finite Turing machine, that is,
finite universal Turing machines are not universal
because, taking any finite universal Turing machine,
there will always be some computations that fail in it,
but that do not fail in other
equivalent finite Turing machines
that have more tape or more time;
more time perhaps because
they are faster or more durable.

On the other hand,
any finite universal Turing machine computation that {\sc halt}s
will be run identically by its corresponding Turing machine,
which is a universal Turing machine.
And this means that, save for time or tape limitations,
every finite universal Turing machine computes exactly
as its corresponding universal Turing machine.

This also means that
every Turing complete device can calculate,
save for time or external memory limitations, 
whatever any Turing machine can compute.
There are actually real Turing complete devices.
For instance, every general-purpose computer {\sc cpu}
is a Turing complete device, by design.

\section{Calculability}

We will assume the following thesis:
\thesis(2){What is effectively calculable is
what a person can calculate.}

This could be denied from a Platonist view of mathematics,
because a super person, or a super machine
with a super syntax engine,
could calculate what a plain person cannot.
But, firstly, we should say that this would be 
“super calculable”, not just “calculable”,
and even less “effectively calculable”.

And, secondly, if those super calculations are
outside our plain syntactic capabilities,
then we could neither identify those super calculations
as calculations,
nor we could follow nor understand them.
For seeing this, just imagine any finished calculation.
If each and every step of the calculation
obeys the finite set of rules for the calculation,
that is, if each and every step obeys the algorithm,
then we have a plain calculation
that we can follow and understand.
Or else, the steps that do not obey
the algorithm are errors,
and then the calculation is wrong,
even if its final result is right.
So, for us, a super calculation can only be
a plain calculation or a wrong calculation.

Another objection could be that different
persons can have different calculating capabilities.
This is true, but to investigate this issue,
let us present our own thesis:
\thesis(3){Persons' syntax engine
 is a finite universal Turing machine.}
A simpler way to express this is to say that
{\it we are Turing complete}.

If this is the case, then, for normal persons,
that is, for persons without mental disabilities,
the difference can only be the amount of memory or time.
As a consequence, given access to enough memory,
everybody has the same calculating capability,
though the amount of time needed to perform
any particular calculation
would not be the same for everyone.

\section{Proof}

We will now show that if thesis 3 is true,
then Church's thesis (thesis 1) is also true. 

First,
we take thesis 1 in \refsc{Church's Thesis} and,
using thesis 2 in \refsc{Calculability},
we replace ‘what is effectively calculable’
with ‘what a person can calculate’.
Second, using the concept of syntax engine,
seen in \refsc{Syntax Engine},
what a person can calculate is what her syntax engine can calculate.
Now, if thesis 3 in \refsc{Calculability} is true, that is,
if persons' syntax engine is a finite universal Turing machine,
then what a person's syntax engine can calculate
is what a finite universal Turing machine can compute.
A finite universal Turing machine is a finite Turing machine,
as seen in \refsc{Finite Turing Machine},
and anything that a finite Turing machine can compute
can also be computed by its corresponding Turing machine,
as shown in \refsc{Effective Computation}.
To close this proof just remember the first definition
in \refsc{Church's Thesis}:
anything that a Turing machine can compute is computable.
{\sc qed}

\section{Discussion}

From the proof it is easy to see that it is also possible
to imply Church's thesis from a weaker assumption:
\thesis(4){Persons' syntax engine is a finite Turing machine.}
But this thesis 4 does not fit some empirical facts
because, if thesis 4 were true, but thesis 3 were false,
then some computations would be outside
the syntactic capability of persons.
This would mean that we persons
could not follow some Turing machine computations.

On the other hand, the stronger thesis 3 implies thesis 4, and it
is very close to the converse of Church's thesis.
The converse of Church's thesis would be:
\thesis(5){What is computable is effectively calculable.}
For this thesis 5 to be true,
a person's syntax engine should
have to be a universal Turing machine,
but then persons could calculate any computation,
that is, any algorithm with any data.
Sadly, we persons are finite, and therefore
we have neither access to an infinite memory,
nor an infinite amount of time to do calculations.

Luckily, if thesis 3 is true,
meaning that we persons are Turing complete,
then we can calculate any computation,
except for limitations of memory or time.
So, taking a small enough part of any computation,
we can calculate whether it is computed rightly or wrongly.

\section{Conclusions}

We can summarize thesis 2 and thesis 3 in a refined Church's thesis:
\thesis(6){What is effectively calculable is effectively computable.}

\label{Thesis6}{Euler Diagram of a Refined Church's Thesis}
\MTbeginfig(170bp, 100bp, 0bp);  
\MT: d := 3bp;
\MT: pickup pencircle scaled 1 bp;
\MT: z0 = (w/2,3h/4 - d);
\MTlabel(0)"Effectively calculable";
\MT: draw (x0l-d,y0b-d) ... (w/2,y0b-3d) ... (x0r+d,y0b-d) ...
\MT:      (x0r+d,y0t+d) ... (w/2,y0t+3d) ... (x0l-d,y0t+d) ... cycle;
\MT: z1 = (w/2, h/2 - 2d);
\MTlabel(1)"Effectively computable";
\MT: draw (x0l-2d,y1b-d) ... (w/2, y1b-3d) ... (x0r+2d,y1b-d) ...
\MT:      (x0r+2d,y0t+2d) ... (w/2, y0t+5d) ... (x0l-2d,y0t+2d) ... cycle;
\MT: z2 = (w/2, h/4 - 4d);
\MTlabel(2)"Computable";
\MT: draw (x0l-3d,y2b-d) ... (w/2, y2b-3d) ... (x0r+2d,y2b-d) ...
\MT:      (x0r+3d,y0t+3d) ... (w/2, y0t+7d) ... (x0l-3d,y0t+3d) ... cycle;
\MT: 
\MT: 
\MTendfig;
\continuepar$$\box\MTbox$$

Save for time or memory limitations,
we can calculate whatever any Turing machine can compute.
In other words, we are Turing complete.
Turing could not have designed his universal machine
if his own syntax engine were not
a finite universal Turing machine.
Turing was Turing complete!

In \refsc{Proof} we have shown that,
if we are Turing complete,
then Church's thesis is true.
By contraposition, this implies that
if Church's thesis were false, then
we would not be Turing complete;
that is, we would not be {\it just\/} Turing complete,
because then we could perform some super calculations
that no Turing machine can compute.

Therefore, the {\it natural law\/} stating
“the limitations of the mathematicizing power
of [our species {\it Homo sapiens\/}]” can be formulated as:
$$\hbox{\it In calculating capability, we are just Turing complete.}%
  \leqno\diamond\ \hbox{Post's Law}$$
This way \cite Post (1936) program is accomplished.

\section{Epilogue on Post (1936) Program}

\cite Church (1937) reviewed
\cite Post (1936) program unfavorably.
For Church, his thesis gives a definition
of effective calculability as computability
that seems “to be an adequate representation
of the ordinary notion”, and that is all.
However, even if that were the case, considering
Church's thesis a definition does not settle
the question, but just sweep it under the carpet.
For suppose two possibilities:
\beginpoints
\point[A)] Tomorrow someone devises a procedure to
   perform calculations that are beyond
   the capability of any Turing machine.
\point[B)] Tomorrow some machine is found
   that performs calculations that are beyond
   the capability of any Turing machine.
\endpoints
In either case, Church's definition would be wrong.

So it is not that effective calculability does not
have an exact definition, and that equating it to computability
solves the problem, as \cite Church (1937) argued.
What happens is that Church's thesis is
uncertain because it depends on
what it might occur tomorrow,
showing that it is empirically refutable.

For \cite Post (1936) program,
effective calculability is defined exactly
(by thesis~2, see \refsc{Calculability}),
and then Church's thesis is not a definition,
but a {\it natural law\/} that
states a limitation of the 
calculating capability of our species {\it Homo sapiens\/}
(our thesis 3, see \refsc{Calculability},
 reformulated in \refsc{Conclusions},
 for which thesis 1, see \refsc{Church's Thesis},
 is a consequence, see \refsc{Proof}).
Therefore, Post's law is our thesis 3
as expressed in \refsc{Conclusions}
and, as any {\it natural law}, it is refutable
and then it is in need of continual verification.
It is refutable because
it predicts both the impossibility of case~A,
and that we are blind to case~B.

As in our argument in favor of thesis 2,
see \refsc{Calculability},
we are blind to case~B if
our calculating capability is limited to computing,
because then we cannot see the super calculation
as a proper calculation.
This is possibly the case of quantum systems.
Take for example a Schrödinger's cat in its box.
If we suppose that the box is a physical device
that performs a calculation the result of which
is a dead or an alive cat,
then we have to conclude that
no algorithm is obeyed, but that
some random activity happens.

\cite Post (1936) program is deep.
As he concludes, in page 105:
\begincitation
Only so, [that is, only if Church's thesis is a
{\it natural law} stating the limitations of 
the mathematicizing power of our species {\it Homo sapiens}],
can Gödel's theorem concerning the incompleteness
 of symbolic logics of a certain general type and
Church's results on the recursive unsolvability
 of certain problems
be transformed into conclusions concerning all symbolic
logics and all methods of solvability.
\endcitation

\vfill\break

\xsection{Figures}\smallskip

\leavevmode\ref{Thesis1}, in \refsc{Thesis1}, page \refpg{Thesis1}.

\leavevmode\ref{Thesis6}, in \refsc{Thesis6}, page \refpg{Thesis6}.

\references
\item Chomsky (1957)
\item Chomsky (1959)
\item Church (1937)
\item Gandy (1980)
\item Hilbert (1922)
\item Post (1936)
\item Turing (1936)
\endreferences

\bye

\xsection{References}
\nobreak\smallskip

\reference Chomsky (1957):
Noam Chomsky, \book{Syntactic Structures};
The Hague, Mouton \& Co., 1957, 2002.
\ISBN 3-11-017279-8.

\reference Chomsky (1959):
Noam Chomsky, \article{On Certain Formal Properties of Grammars};
in \periodical{Information and Control},
Volume 2, Issue 2, pp.~137--167, June 1959.\\
\DOI{10.1016/S0019-9958(59)90362-6}.

\reference Church (1937):
Alonzo Church, \article{Review of \cite Post (1936)};
in \periodical{The Journal of Symbolic Logic},
Volume 2, Issue 1, p.~43, March 1937.
\DOI{10.1017/S0022481200039591}.

\reference Gandy (1980):
Robin Gandy,
\article{Church's Thesis and Principles for Mechanisms};
\DOI{10.1016/s0049-237x(08)71257-6}.
In \book{The Kleene Symposium}
(editors: J.\ Barwise, H.J.\ Keisler \& K.\ Kunen),
Volume 101 of
Studies in Logic and the Foundations of Mathematics;
North-Holland, Amsterdam, 1980, pp.~123--148;
\ISBN 0-444-85345-6.

\reference Hilbert (1922):
David Hilbert,
„Neubegründung der Mathematik: Erste Mitteilung“;
in \periodical{Abhandlungen aus dem Seminar
 der Hamburgischen Universität},
Volume 1, Issue 1, pp.~157--177, December 1922;
\DOI{10.1007/bf02940589}.
Spanish translation by L.F.\ Segura as
\article{La Nueva Fundamentación de las Matemáticas},
in \book{Fundamentos de las Matemáticas},
 (editors: C.\ Álvarez \& L.F.\ Segura),
Servicios Editoriales de la Facultad de Ciencias, UNAM,
México, 1993, pp.~37--62;
\ISBN 968-36-3275-0.
English translation by W.B.\ Ewald as
\article{The New Grounding of Mathematics: First Report},
in \book{From Kant to Hilbert:
 A Source Book in the Foundations of Mathematics}, Volume 2,
 (editor: W.B.\ Ewald),
Oxford University Press, Oxford, 1996, pp.~1115--1134;
\ISBN 0-19-850536-1.

\reference Post (1936):
Emil Post,
\article{Finite Combinatory Processes --- Formulation 1};
in \periodical{The Journal of Symbolic Logic},
Volume\ 1, Number\ 3, pp.~103--105, September 1936.
Received October 7, 1936.
\DOI{10.2307/2269031}.

\reference Turing (1936):
Alan Turing, \article{On Computable Numbers,
 with an Application to the Entscheidungsproblem};
in \periodical{Proceedings of the London Mathematical Society},
Volume s2-42, Issue 1, pp.~230--265, 1937.
Received 28 May, 1936. Read 12 November, 1936.
\DOI{10.1112/plms/s2-42.1.230}.

\bye